\documentclass[%
reprint,
superscriptaddress,
showpacs,
preprintnumbers,
 amsmath,amssymb,
 aps,
prb
]{revtex4-1}

\usepackage{color}

\usepackage{graphicx}% Include figure files
\usepackage{dcolumn}% Align table columns on decimal point
\usepackage{multirow}
\usepackage{bm}% bold math
\usepackage{url}

\newcommand{\blue}[1]{{\color{black}#1}}
\newcommand{\red}[1]{{\color{black}#1}}
\usepackage[normalem]{ulem}

\begin{document}


%\preprint{APS/123-QED}

\title{Monte Carlo analysis for finite temperature magnetism of Nd$_2$Fe$_{14}$B permanent magnet}% Force line breaks with \\
%influence of the quantum fluctuations 




\author{Yuta~\surname{Toga}}
\affiliation{ESICMM, National Institute for Materials Science (NIMS), Tsukuba, Ibaraki 305-0047, Japan}

\author{Munehisa~\surname{Matsumoto}}
\affiliation{ESICMM, National Institute for Materials Science (NIMS), Tsukuba, Ibaraki 305-0047, Japan}
\author{Seiji~\surname{Miyashita}}
\affiliation{Department of Physics,~The University of Tokyo, Bunkyo-Ku 113-0033, Japan}
\affiliation{ESICMM, National Institute for Materials Science (NIMS), Tsukuba, Ibaraki 305-0047, Japan}

\author{Hisazumi~\surname{Akai}}
\affiliation{Institute for Solid State Physics,~The University of Tokyo, Kashiwa 277-8581, Japan}
\affiliation{ESICMM, National Institute for Materials Science (NIMS), Tsukuba, Ibaraki 305-0047, Japan}

\author{Shotaro~\surname{Doi}}
\affiliation{Institute for Solid State Physics,~The University of Tokyo, Kashiwa 277-8581, Japan}
\affiliation{ESICMM, National Institute for Materials Science (NIMS), Tsukuba, Ibaraki 305-0047, Japan}

\author{Takashi~\surname{Miyake}}
\affiliation{CD-FMat, National Institute of Advanced Industrial Science and Technology (AIST), Tsukuba, Ibaraki 305-8568, Japan}
\affiliation{ESICMM, National Institute for Materials Science (NIMS), Tsukuba, Ibaraki 305-0047, Japan}
\author{Akimasa~\surname{Sakuma}}
\affiliation{Department~of~Applied Physics, Tohoku~University, Sendai 980-8579, Japan}
\affiliation{CREST, Japan Science and Technology Agency (JST), Chiyoda, Tokyo 102-0075, Japan}

%\thanks{}

%option in \documentclass). \noaffiliation ris required (may also be

\date{\today}
%%%%%%%%%%%%%%%%%%%%%%%%%%%%%%%%%%%%%

\begin{abstract}
%Spin-1 Bose-Hubbard model is expected to realize in the cold atom physics.

%%%%%%%%%%%%%%%%%%%%%%%%%%%%%%%%%%%%%%%%%%%%%%%%/

We investigate the effects of magnetic inhomogeneities and thermal fluctuations on the magnetic properties of a rare earth intermetallic compound, Nd$_2$Fe$_{14}$B.
The constrained Monte Carlo method is applied to a Nd$_2$Fe$_{14}$B bulk system to realize the experimentally observed spin reorientation and magnetic anisotropy constants $K^{\rm A}_m (m=1, 2, 4)$ at finite temperatures.
Subsequently, it is found that the temperature dependence of $K^{\rm A}_1$ deviates from the Callen--Callen law, $K^{\rm A}_1(T) \propto M(T)^3$, even above room temperature, $T_{\rm R}\sim 300\rm\, K$, when the Fe (Nd) anisotropy terms are removed to leave only the Nd (Fe) anisotropy terms.
This is because the exchange couplings between Nd moments and Fe spins are much smaller than those between Fe spins.
It is also found that the exponent $n$ in the external magnetic field $H_{\rm ext}$ response of barrier height $\mathcal{F}_{\rm B}=\mathcal{F}_{\rm B}^0(1-H_{\rm ext}/H_0)^n$ is less than $2$ in the low-temperature region below $T_{\rm R}$, whereas $n$ approaches $2$ when $T>T_{\rm R}$, indicating the presence of Stoner--Wohlfarth-type magnetization rotation.
This reflects the fact that the magnetic anisotropy is mainly governed by the $K^{\rm A}_1$ term in the $T>T_{\rm R}$ region.

%%%%%%%%% %%%%%%%%%%%%%%%%%%%%%%%%%%%%%%%%%%%%%%/ 

\end{abstract}

%\keywords{}

\maketitle

%\section{introduction}
%\subsection{}
%\subsubsection{}

%%%%%%%%%%%%%%%%%%%%%%%%%%%%%%%%%%%%%%
%%%%%***************** Introduction *******************
%%%%%%%%%%%%%%%%%%%%%%%%%%%%%%%%%%%%%%

\section{Introduction}

Rare earth permanent magnets, particularly Nd-Fe-B, exhibiting strong magnetic performance\cite{herbst_ndfe_14b_1991} are attracting considerable attention because of the rapidly growing interest in electric vehicles.
The main focus of research in involving these materials is to increase the coercive field $H_c$ and improve the temperature dependence.\cite{fidler_electron_1989,vial_improvement_2002,li_effect_2009,woodcock_understanding_2012,hono_strategy_2012}
Therefore, a number of studies have conducted micromagnetic simulations \cite{kronmuller_angular_1987,sakuma_micromagnetic_1990,fischer_grain-size_1996,hrkac_role_2010} for the magnetization processes using inhomogeneous magnetic parameters to describe the complex structures in sintered magnets.
Many of the results predict that the distinctive feature of magnetic anisotropy near the grain boundaries of Nd-Fe-B particles is responsible for the degradation of $H_c$.

Thus, one of the remaining subjects of the theoretical study is to give quantitative aspects in microscopic viewpoint or in atomic-scale, to the $m$-th order magnetic anisotropy constants $K^{\rm A}_m$ and their temperature dependence near the grain surfaces or grain boundaries.
For $K^{\rm A}_1$ at the surface of Nd-Fe-B particles, Moriya et al. \cite{moriya_first_2009} and Tanaka et al. \cite{tanaka_first_2011} calculated the crystal field parameter $A_2^0$ using a first-principles technique and pointed out that $K^{\rm A}_1$ (mainly proportional to $A_2^0$) is negative at the $(001)$ surface when the $(001)$ Nd layer is exposed to a vacuum.
However, few theoretical studies have examined the temperature dependence of $K^{\rm A}_m$, even for the bulk system,
since the qualitative theory was developed by Callen and Callen.\cite{callen_static_1963,callen_present_1966,skomski_permanent_1999}%
Recently, Sasaki et al. \cite{sasaki_theoretical_2015} and Miura et al. \cite{miura_direct_2015} conducted theoretical studies in the quantitative level on the temperature dependence of $K^{\rm A}_m$ for a Nd$_2$Fe$_{14}$B bulk system based on crystal field theory, and successfully reproduced various experimental results.
However, as these theories relied on the mean field approach in terms of the exchange coupling between the Nd $4f$ moments and Fe $3d$ spins, the results cannot be directly applied to $K^{\rm A}_m$ near the surfaces or interfaces of particles.
Moreover, because the crystal field analysis employed in these works is based on a quantum mechanical approach, which is typical for $4f$ electronic systems,\cite{yamada_crystal-field_1988} it is effectively impossible to treat finite systems of nm- or $\mu$m-scale using a similar method.

Therefore, in the present work, in anticipation of future work on magnetization reversal in finite-sized particles, we employed a realistic model with a classical Heisenberg Hamiltonian to calculate the magnetic properties of a Nd$_2$Fe$_{14}$B bulk system at finite temperatures.
The key features of our model are:
1) an appropriate crystalline electric field Hamiltonian\cite{yamada_crystal-field_1988} is included in the classical manner,
2) exchange coupling parameters are obtained by first-principles calculations,
3) $K^{\rm A}_m$ is directly evaluated from Monte Carlo (MC) simulations without employing the mean field analysis, and
4) the constrained Monte Carlo (C-MC) method,\cite{asselin_constrained_2010} is adopted to evaluate the temperature dependence of magnetic anisotropy.
Note that we can naturally realize the experimentally observed spin reorientation and $K^{\rm A}_m$.
Reflecting the (inhomogeneous) variation of magnetic parameters in the unit cell composed of $68$ atoms (see Fig.~\ref{fig:ndfeb}), $K^{\rm A}_1$ does not obey the Callen--Callen law,\cite{callen_static_1963,callen_present_1966} which states that $K^{\rm A}_1(T)\propto M(T)^3$ when considering only the Nd (Fe) anisotropy terms and neglecting the Fe (Nd) anisotropy terms.
We also analyze the response of the external magnetic field $H_{\rm ext}$\cite{gaunt_magnetic_1986,chantrell_calculations_2000,suess_reliability_2007, bance_thermally_2015,bance_thermal_2015,goto_energy_2015}
for a barrier height $\mathcal{F}_{\rm B}(H_{\rm ext})=\mathcal{F}_{\rm B}^0(1-H_{\rm ext}/H_0)^n$, and
find that the $H_{\rm ext}$ response deviates from the Stoner--Wohlfarth-type ($n=2$), especially below room temperature, $T_{\rm R}\sim 300\,\rm K$.
%
%This paper is organized as follows.
%In Sec.~II, we present our model and explain the method.  In Sec.~III, the results of our calculations are presented.

%%%%%%%%%%%%%%%%%%%%%%%%%%%%%%%%%%%%%%
%%%%%***************** Formulation *******************hat{J}_i^
%%%%%%%%%%%%%%%%%%%%%%%%%%%%%%%%%%%%%%

\section{Model and Method}
\subsection{Model}
%
%%%%%%%%%%%%%%%%%%%%%%%%%%%%%%%%%%%%%%%%%%%%%%%%%%%%%%
%%%%%%     fig.1   Structure of Nd-Fe-B %%%%%
\begin{figure}[t]
\includegraphics[width=6cm]{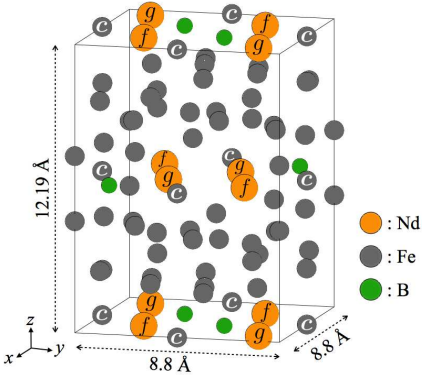}
\caption{\label{fig:ndfeb}
Unit cell of Nd$_2$Fe$_{14}$B including $68$ atoms (space group:$P4_2/mnm$  (No.136)). \cite{herbsT_Relationships_1984}
Only Nd($f, g$) and Fe$(c)$ sites are represented.
This figure was plotted using VESTA.\cite{momma_vesta_2011}}
\end{figure}
%%%%%%%%%%%%%%%%%%%%%%%%%%%
%
%We consider the three dimensional classical Heisenberg Hamiltonian as follow:
By treating each atom as having classical spin, we constructed a three-dimensional Heisenberg model including realistic atom locations for Nd$_2$Fe$_{14}$B, as shown in Fig.~\ref{fig:ndfeb}.
This model using atomic-scale parameters was defined as follow:
%
%We constructed classical Heisenberg models by using parameters from first principle calculations,
%and analyzed these models based on Monte Carlo simulations.
%
%
%----------------------------------------------------------------------
\begin{eqnarray}
 {\cal H} &=& -2\sum_{i<j} S_i J^{\rm ex}_{ij} S_j {\bm{e}}_i \cdot {\bm{e}}_j
 		      -\mu_0 \sum_i  m_i \bm{e}_i \cdot \bm{H}_{\rm ext} \nonumber \\
 		  & & -\sum_{i \in {\rm TM} } D_{i}^{\rm A} (e^z_i)^2 \nonumber 
 		  +\sum_{i \in {\rm RE}} \sum_{l=2,4,6} \tilde{\theta}_l^{{J}_i} A_{l,i}^{m_l}\langle r^l\rangle_i \hat{\mathcal{O}}^{m_l}_{l,i},\\
% 	      +\sum_{i \in {\rm RE}} \sum_{l=2,4,6} B_{l,i}^{m_l=0} \hat{\mathcal{O}}^{m_l=0}_{l,i},
 	     \label{eq:hami}
\end{eqnarray}
%----------------------------------------------------------------------
%
where
$S_iJ^{\rm ex}_{ij}S_j$ is the exchange coupling constant including the spin amplitude between the $i$-th and $j$-th sites,
${\bm e}_i$ is the normalized spin vector at the $i$-th site,
$m_i$ is the magnetic moment, $\mu_0$ is the magnetic permeability of a vacuum and $\bm{H}_{\rm ext}$ is the external magnetic field.
The third and fourth terms include single-ion magnetic anisotropy properties. We consider transition metals (TM) and rare-earth elements (RE) separately.
The anisotropy of TM sites is defined using the magnetic anisotropy parameter $D^{\rm A}_i$ and the $z$-component of $\bm{e}_i$, i.e., $e^z_i$.
%
% ${\theta}_{i}$ is angle of $\bm{{e}}_{i}$ from $z$-axis.(J^z_i)
%
%$B_{l,i}^{m_l}= \theta_l^{{J}_i} A_{l,i}^{m_l}\langle r^l\rangle_i$ is crystalline electronic field coefficient and $\hat{\mathcal{O}}_{l,i}^{m_l}$ is Stevens operator.
The anisotropy of RE sites is based on crystal field theory\cite{stevens_matrix_1952,yamada_crystal-field_1988} and uses the Stevens operator $\hat{\mathcal{O}}_{l,i}^{m_l}$, crystal field parameter $A_{l,i}^{m_l}$, and Stevens factor $\tilde{\theta}_l^{{J}_i}$.
Here, $\langle r^l \rangle_i$ can be calculated as the spatial average of the $4f$ electron distribution.
In the present paper, we consider $m_l=0$ for simplicity.
For reference, note that $\hat{\mathcal{O}}_{l,i}^{m_l=0}$ and $\tilde{\theta}_l^{J=\frac{9}{2}}$:
% 
%----------------------------------------------------------------------
\begin{eqnarray}
	\hat{\mathcal{O}}_{2,i}^0&=&3(J^z_i)^2-J_i^2, \nonumber \\
	\hat{\mathcal{O}}_{4,i}^0&=&35(J^z_i)^4-\bigl[30{J}_i^2-25\bigr](J^z_i)^2+\bigl[3{J}_i^4-6{J}_i^2\bigr],  \nonumber \\
	\hat{\mathcal{O}}_{6,i}^0&=&231(J^z_i)^6-\bigl[ 315{J}_i^2-735 \bigr](J^z_i)^4 \nonumber\\
		 && \quad \qquad + \bigl[105{J}_i^4-525{J}_i^2+294 \bigr](J^z_i)^2  \nonumber \\
		 && \quad \qquad - \bigl[ 5{J}_i^6-40{J}_i^4+60{J}_i^2 \bigr], \label{eq:stevens}
\end{eqnarray}
\begin{eqnarray*}
	\tilde{\theta}_2^{\rm \frac{9}{2}}= \frac{-7}{3^2\cdot 11^2},\ 
	\tilde{\theta}_4^{\rm \frac{9}{2}}= \frac{-2^3\cdot 17}{3^3\cdot 11^3 \cdot 13},\ 
	\tilde{\theta}_6^{\rm \frac{9}{2}}= \frac{-5\cdot 17 \cdot 19}{3^3\cdot 7\cdot 11^3 \cdot 13^2}, \nonumber
\end{eqnarray*}
%
%----------------------------------------------------------------------
%	$(J^z_i)&=&{J}_i {e}^z_j$
where $J^z_i={J}_i {e}^z_i$ is the $z$-component of the total angular momentum ${J}_i$,
which is $9/2$ for Nd atoms, and we use $J_i^2$ instead of $J_i(J_i+1)$ in the classical manner.	
%Now, we assume that the length of $\bm{J}=J\bm{e}$ is not $\sqrt{ J(J+1) }$ but $J$ as classical spin.

%%%%%%%%%%%%%%%%%%%%%
\begin{table}[b]
\caption{\label{tab:parameter} Site occupancies and model parameters of each crystallographically inequivalent atom.
The spin magnetic moments, $m^s$, are calculated from the first-principles calculation code, Machikaneyama (AkaiKKR).\cite{akaikkrmachikaneyama}
The anisotropy parameters $D^{\rm A}_{i_{\rm TM}}$ and $A_l^{\red{m_l}}\langle r^l \rangle $ are taken from previous results.\cite{miura_magnetocrystalline_2014,yamada_crystal-field_1988}
We neglected the $D^{\rm A}$ values of B and Nd, as they are less than $0.1\,\rm meV$,
and used the $\langle r^l \rangle$  values of Nd, Ref.~\citenum{freeman_theoretical_1962}, i.e., $\langle r^2 \rangle=1.001\,\red{a_B^2}$, $\langle r^4 \rangle=2.401\,\red{a_B^4}$, and $\langle r^6 \rangle=12.396\,\red{a_B^6}$\red{, where $a_B$ is the Bohr radius}.
}
\begin{ruledtabular} 
\begin{tabular}{lcccccc}
atom &occ.&$m^s$ [$\mu_B$] & $D^{\rm A}_{i_{\rm TM}}$~[meV] & \multicolumn{3}{c}{$A_l^{\red{m_l}}\langle r^l \rangle $~[K]} \\
\hline
B($g$)      &4& -0.169 & -      & \multicolumn{3}{c}{-} \\
Fe($c$)     &4&  2.531 & -2.14  & \multicolumn{3}{c}{-} \\
Fe($e$)     &4&  1.874 & -0.03  & \multicolumn{3}{c}{-} \\
Fe($j_1$)   &8&  2.298 &  1.07  & \multicolumn{3}{c}{-} \\
Fe($j_2$)   &8&  2.629 &  0.58  & \multicolumn{3}{c}{-} \\
Fe($k_1$)  &16&  2.063 &  0.55  & \multicolumn{3}{c}{-} \\
Fe($k_2$)  &16&  2.206 &  0.38  & \multicolumn{3}{c}{-} \\ \hline
        & &      &\multicolumn{1}{r}{($l,\red{m_l}$):} & ($2, 0$)&($4, 0$)&($6, 0$)\\\hline
Nd($f$)     &4& -0.413 & -      & \multirow{2}{*}{295.3}& \multirow{2}{*}{-29.5}& \multirow{2}{*}{-22.8} \\
Nd($g$)     &4& -0.402 & -      & \\
\end{tabular}
\end{ruledtabular}

\end{table}
%%%%%%%%%%%%%%%%%%%%%%%%%%%
%
Table \ref{tab:parameter} lists the atomic-scale parameters used in the present study.
%The tetragonal unit cell of Nd$_2$Fe$_{14}$B consists of 68 atoms, which occupy nine crystallographically inequivalent sites as seen in Table \ref{tab:parameter}.
The $68$ atoms in the tetragonal unit cell of Nd$_2$Fe$_{14}$B (see Fig.~\ref{fig:ndfeb}) occupy nine crystallographically inequivalent sites, as seen in Table \ref{tab:parameter}.
These atom locations and lattice constants ($a=b=8.8$\,\AA, $c=12.19$\,\AA) were set to experimental values.\cite{herbsT_Relationships_1984}
$m^s$ is the spin magnetic moment of valence electrons (excluding $4f$-electrons).
We defined $m_i=m^s_i$ for Fe and B atoms, and $m_i=m^s_i+m^{4f}_i$ for Nd atoms.
Here, the magnetic moment of $4f$-electrons in each Nd atom is $m^{4f}=8J/11\,\mu_B \sim 3.273\,\mu_B$.
For the magnetic anisotropy terms,
%$D^{\rm A}_{i_{\rm TM}}$ is adopted the results\cite{miura_magnetocrystalline_2014} of first principles calculation for Y$_2$Fe$_{14}$B which have similar crystal structure with Nd$_2$Fe$_{14}$B.
the $D^{\rm A}$ values were set to previous first-principles calculation results\cite{miura_magnetocrystalline_2014} for Y$_2$Fe$_{14}$B, which has a similar crystal structure as Nd$_2$Fe$_{14}$B.
In contrast, we adopted experimental results\cite{yamada_crystal-field_1988} regarding $A_l^{\red{m_l}}$,
even though some research for the $A_l^{\red{m_l}}$ values of Nd$_2$Fe$_{14}$B was performed using first-principles calculations.\cite{hummler_full-potential_1996,yoshioka_crystal_2015}
This is because first-principles evaluations of $A_l^{\red{m_l}}$ are strongly dependent on the calculation conditions;
in particular, the values of the $l=6$ terms are still open to some debate.
%especially, values of $l=6$ terms are still arguable. 
%, there is still room for discussion 

%%%%%%     fig. Blm
\begin{figure}%[b]
\includegraphics[width=8cm]{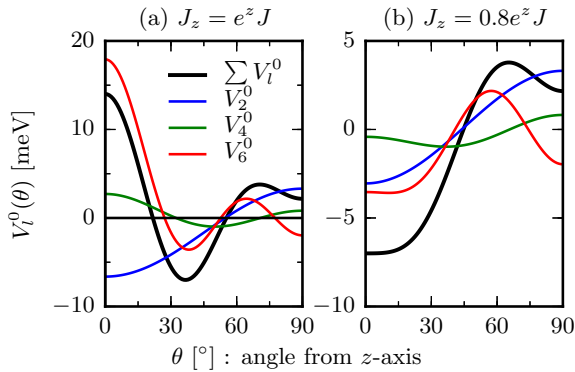} % Here is how to import EPS art
\caption{\label{fig:blm} Anisotropy potentials $V_l^0(\theta)\,\red{\rm [meV]}$ for $J=9/2$ single spin. Black lines denote total anisotropy potential $V_2^0+V_4^0+V_6^0$.
}
\end{figure}
%%%%%%%%%%%%%%%%%%%%%%%%%%%
%
The higher-order crystal field parameters $A_4^0$ and $A_6^0$ of the Nd atoms have a significant effect on the low-temperature properties of Nd$_2$Fe$_{14}$B.
%Nd$_2$Fe$_{14}$B has much effect of $lA_4^0$ and $A_6^0$ at low temperature.
%
%To see these effects clearly, 
To illustrate these effects, Fig.~\ref{fig:blm} shows the anisotropy potential for $J=9/2$ single classical spin:
\begin{eqnarray}
V_l^{m_l}(\theta) = \tilde{\theta}_l^\frac{9}{2} A_{l}^{m_l}\langle r^l\rangle \hat{\mathcal{O}}^{m_l}_{l}(\theta),
\end{eqnarray}
where $\theta$ is the spin angle measured from the $z$-axis (i.e., $e^z=\cos\theta$) and $A_{l}^{m_l}\langle r^l\rangle$ take the values in Table~\ref{tab:parameter}.
The potential $V_2^0$ increases monotonically as $\theta$ increases, whereas $V_4^0$ and $V_6^0$ vary non-monotonically.
Because of this behavior,
the total anisotropy potential $V_2^0+V_4^0+V_6^0$ attains a minimum at $\theta=36.7^\circ$ for (a) $J_z=e^zJ$.
% which is corresponding to $T=0$.
In contrast, for (b) $J_z=0.8e^zJ$, the minimum occurs at $\theta=0^\circ$.
This coefficient ($=0.8$) of $e^z$ can be regarded as an effect of thermal fluctuations at $T>0$.
The above results indicate that the spin direction is tilted from the $z$-axis at $T=0$,
although this tilting disappears at a certain temperature. This behavior corresponds to the spin reorientation phenomenon.
In the case of Nd$_2$Fe$_{14}$B, the spin reorientation transition is due to the $V_4^0$ and $V_6^0$ values of Nd atoms, and includes the effects of exchange couplings and the magnetic anisotropy of Fe atoms (for details, see Sec.~\ref{sec:thermo}).

%spin reorientation transition was observed in experimental.

%
%%%%%%%%%%%%%%%%%%%%%%%%%%%%%%%%%%%%%%%%%%%%%%%%%%%%%%
%%%%%%     fig.   exchange coupling %%%%%
\begin{figure}%[b]
\includegraphics[width=8cm]{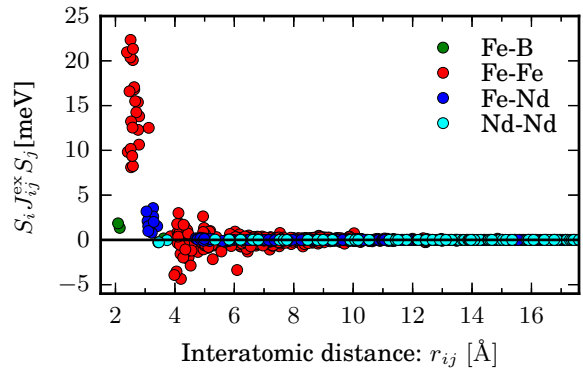}
\caption{\label{fig:jex} Exchange coupling constant between each atom as a function of interatomic distance.}
\end{figure}
%%%%%%%%%%%%%%%%%%%%%%%%%%%
%
Figure \ref{fig:jex} shows the exchange coupling constants, $S_i J^{\rm ex}_{ij}S_j$, which include the spin amplitude as a function of interatomic distance $r_{ij}$.
These constants were calculated with Liechtenstein's formula \cite{liechtenstein_local_1987}
that has been implemented on the first-principles electronic-structure calculation using the Korringa-Kohn-Rostoker (KKR) Green's function method,
Machikaneyama (AkaiKKR).\cite{akaikkrmachikaneyama}
In the calculation, standard muffin-tin-type potentials were assumed, and
the local density approximation parametrized by Morruzi, Janak and Williams \cite{first-principles} was used.
Up to $d$-wave scatterings were taken into account in KKR,
and ($8\times 8\times 6$) $k$-points in the first Brillouin zone being used for the calculation of $J^{\rm ex}_{ij}$'s.
For the Nd 4{\it f}-states, the so called open-core approximation was employed.

%These constants are calculated with Lichtenstein's formula\cite{liechtenstein_local_1987}
%has been implemented on the first principles electronic-structure calculation code, Machikaneyama (AkaiKKR)\cite{akaikkrmachikaneyama}.
%%
% calculation condition...
%

%
%
From Fig.~\ref{fig:jex}, we can see that the exchange couplings between Fe and Nd have much smaller values than those between Fe atoms.
In addition, none of the Nd atoms interacts directly with other Nd atoms.
The \red{amplitude} relation of the exchange couplings is consistent with
experimental results\cite{herbst_ndfe_14b_1991} based on a mean field analysis.
Note that all $S_{\rm Fe} J_{\rm Fe \mathchar"712D Nd}^{\rm ex} S_{\rm Nd}$ on $r_{ij} < 4$\,{\AA} have positive values in Fig.~\ref{fig:jex}. 
As $J_{ij}^{\rm ex}$ is evaluated as the interaction between valence electrons,
$S_{\rm Fe (Nd)}$ can be regarded as being proportional to $m^s_{\rm{Fe (Nd)}}$, i.e., $S_{\rm Nd} S_{\rm Fe} < 0$.
Hence, the bare exchange couplings $J_{\rm Fe \mathchar"712D Nd}^{\rm ex}$ have negative values.
The couplings between Fe and B, $J_{\rm Fe \mathchar"712D B}^{\rm ex}$, also take negative values which can be explained in the same way.

%%%%%%%%%%%%%%%%%%%%%%%%%%%%%%%%%%%%%%%%%%%%%%%%%%%%%%%%%%%%%%%%
%%%%%%%%%%%%%%%%%%%%%%%%%%%%%%%%%%%%%%%%%%%%%%%%%%%%%%%%%%%%%%%%
%%%%%%%%%%%%%%%%%%%%%%%%%%%%%%%%%%%%%%%%%%%%%%%%%%%%%%%%%%%%%%%%
%%%%%%%%%%%%%%%%%%%%%%%%%%%%%%%%%%%%%%%%%%%%%%%%%%%%%%%%%%%%%%%%
%
%
\subsection{Method}
To analyze the finite-temperature magnetism of Nd$_2$Fe$_{14}$B,
we applied MC methods based on the Metropolis algorithm\cite{landau2014guide} to the above classical Heisenberg model.
Although the magnetic anisotropy is evaluated as the magnetization angle dependence of free energy,
this is generally difficult to simulate explicitly using a typical MC approach.
% it is hard to perform explicitly on the basis of typical MC simulations.
%
Therefore, we also adopted the C-MC method\cite{asselin_constrained_2010} to evaluate the magnetic anisotropy.
The C-MC method fixes the direction of total magnetization $\bm{M}=(M_x,M_y,M_z)=(1/N_s)\sum_i m_i \bm{e}_i$ ($N_s$ is the total number of sites) in any direction for each MC sampling without $\bm{H}_{\rm ext}$, 
and then calculates the fixed angle $\theta$ dependencies of free energy $\Delta\mathcal{F}(\theta)$ and magnetization torque $\vec{\mathcal{T}}(\theta)$ as follows:\cite{asselin_constrained_2010}
%
%\begin{eqnarray}
%	P=\min \left[1, \frac{M^\prime_z}{M_z} \frac{|S_{jz}|}{|S^\prime_{jz}|} {\rm e}^{-\beta\Delta\mathcal{H}}  \right]
%\end{eqnarray}
%
%
\begin{eqnarray}
\vec{\mathcal{T}}(\theta)&=& -\left\langle  \sum_i {\bm e}_i \times \frac{\partial\mathcal{H}}{\partial {\bm e}_i }  \right\rangle
\quad \text{for}\ \bm{M}=\bm{M}(\theta), \\
\Delta\mathcal{F}(\theta) &=& \mathcal{F}(\theta) - \mathcal{F}(\theta_0) \nonumber \\
&=& \int_{\theta_0}^\theta d\theta^\prime\, \left[ \red{{\bm n} } (\theta^\prime) \times \vec{\mathcal{T}}(\theta^\prime) \right] \cdot \left.\frac{\partial \red{\bm{n} }(\theta)}{\partial \theta}\right|_{\theta=\theta^\prime},
\end{eqnarray}
where \red{$\bm{n}(\theta)=\bm{M}(\theta)/|\bm{M}(\theta)|$ and} $\bm{M}(\theta)$ is the total magnetization in the fixed direction $\theta$.

Note that Asselin et al.\cite{asselin_constrained_2010} formulated the C-MC method for systems with homogeneous magnetic moments\red{, i.e. all the magnetic moments have same value.}
However, it can easily be extended to \red{systems with inhomogeneous magnetic moments} such as Nd$_2$Fe$_{14}$B.
\red{We now briefly explain only the procedure of the extended C-MC method} \blue{with a fixed $M$ in the direction of $z$-axis:}

\begin{itemize}
\red{
\item[(A)] Select a site $i$ and obtain the new state of $i$-spin \blue{randomely chosen},
\begin{eqnarray*}
\bm{e}_i \rightarrow \bm{e}^\prime_i.
\end{eqnarray*}
\item[(B)] Select a site $j(\neq i)$ randomly.

\item[(C)] Adjust the new state of \blue{the} $j$-spin to preserve $\bm{M}$ direction (namely, $M_x=M_y=0$):
\begin{eqnarray*}
\bm{e}_j &\rightarrow & \bm{e}^\prime_j,\\
&e^{x\prime}_j&= e^{x}_j + \frac{m_i}{m_j}(e^{x}_i - e^{x\prime}_i),\\
&e^{y\prime}_j&= e^{y}_j + \frac{m_i}{m_j}(e^{y}_i - e^{y\prime}_i),\\
&e^{z\prime}_j&= {\rm sign}(e^z_j)\sqrt{ 1- (e^{x\prime}_j)^2 - (e^{y\prime}_j)^2 }.
\end{eqnarray*}
If $1- (e^{x\prime}_j)^2 - (e^{y\prime}_j)^2<0$, return to (A).
\item[(D)] Calculate the new total magnetization:
\begin{eqnarray*}
\bm{M}^\prime &=&\bm{M}+ \frac{1}{N_s}\left[m_i(\bm{e}^\prime_i-\bm{e}_i)+ m_j(\bm{e}^\prime_j-\bm{e}_j)\right].
\end{eqnarray*}
If $\bm{M}^\prime<0$, return to (A).

\item[(E)] Update from the initial spin states ($\bm{e}_i, \bm{e}_j$) to the new spin states ($\bm{e}^\prime_i, \bm{e}^\prime_j$) with \blue{the} probability:
\begin{eqnarray*}
P&=&{\rm min}\left[ 1,  \left(\frac{M_z^\prime}{M_z}\right)^2 \frac{|e_j^z|}{|e_j^{z\prime}|}\exp{(-\beta\Delta E)} \right],\\
\end{eqnarray*}
where \blue{$\beta$ is the inverse temperature and} $\Delta E=E(\bm{e}^\prime_i, \bm{e}^\prime_j)-E(\bm{e}_i, \bm{e}_j)$ is the energy difference.

\item[(F)] Return to (A).
}
\end{itemize}
\red{
To apply C-MC method to the Nd$_2$Fe$_{14}$B bulk system, \blue{we change the procedures (C) and (D) to treat different magnetic moments from those in the original pepar.\cite{asselin_constrained_2010}}
%changed only (C) and (D) from the original paper.
}
%In this formulation, global spin rotation is necessary in order to calculate physical quantities with the arbitrary angle of ${\bm M}$.

The MC (C-MC) simulations in the present study repeated each calculation for 200,000 (100,000) MC steps,
where one MC step is defined as one trial for each spin to be updated.
The first 100,000 (30,000) MC steps were used for equilibration, and the following 100,000 (70,000) MC steps were used to measure the physical quantities.
We performed simulations for 12 different runs with different initial conditions and random sequences.
We then calculated the average results and statistical errors.
To check the system-size dependence,
we used systems of $N_s$=$L^3 \times$ $68$ (unit cell) sites with $L=3$--$6$, imposing the periodic boundary conditions.

%%%%%%%%%%%%%%%%%%%%%%%%%%%%%%%%%%%%%%
%%%%%***************** Results *******************
%%%%%%%%%%%%%%%%%%%%%%%%%%%%%%%%%%%%%%
\section{Results and Discussion}

\subsection{Thermodynamic Properties}\label{sec:thermo}

First, we focus on the magnetic transition points to verify the model and parameter values.
The results in this subsection are based on typical MC, rather than C-MC.
%
%%%%%%     fig. magnetization curve
\begin{figure}%[b]
\includegraphics[width=8cm]{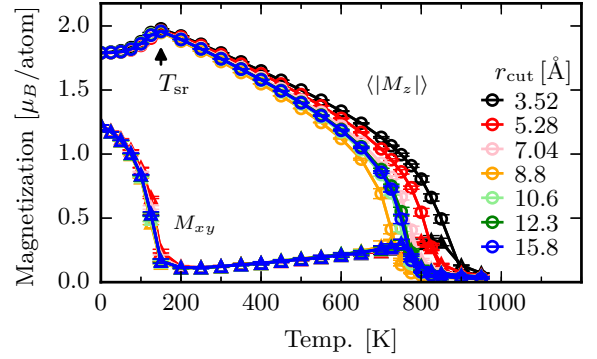} % Here is how to import EPS art

\caption{\label{fig:magdis} Magnetizations as a function of temperature for each effective exchange coupling radius $r_{\rm cut}$. System size is $L=6$.
}
\end{figure}
%%%%%%%%%%%%%%%%%%%%%%%%%%%
%
Figure~\ref{fig:magdis} shows the magnetization curves for each cutoff range $r_{\rm cut}$. We consider all exchange couplings $J_{ij}^{\rm ex}$ under $r_{ij} \leq r_{\rm cut}$.
Here, $\langle A \rangle$ is defined as the statistical average of $A$.
It can be seen that there are two transition points in Fig.~\ref{fig:magdis}.
%

%%%%%%%%%%%%%%%%%%%%%%%%%%%%%%%%%%%%%%%%%
%%%%%%     fig. Tc for each models
\begin{figure}%[b]
\includegraphics[width=8cm]{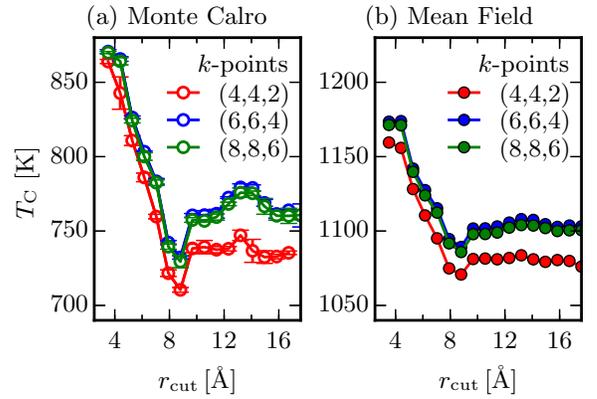} 
\caption{\label{fig:tc} Curie temperatures with (a) MC and (b) mean field as a function of effective exchange coupling radius $r_{\rm cut}$ for the number of $k$-points.
}
\end{figure}
%%%%%%%%%%%%%%%%%%%%%%%%%%%
%
In the higher-temperature region, $\langle |M_z| \rangle$ approaches $0$ at the Curie temperature $T_{\rm C}$.
The magnetization curves show that $T_{\rm C}$ is strongly dependent on $r_{\rm cut}$, even in long-range ($r_{ij} > 3.52\,{\rm \AA} $).
Thus, $T_{\rm C}$ was evaluated more accurately using the Binder parameter\cite{binder_critical_1981,binder_finite_1981,landau2014guide} defined as $g_L=1 - \langle |\bm{M}| \rangle^4 / 3 \langle |\bm{M}|^2 \rangle^2$, for system sizes $L=3$--$6$.
The results are plotted in Fig.~\ref{fig:tc}(a). 
%a
It is apparent that $T_{\rm C}$ has quite different values depending on $r_{\rm cut}$, and the condition of $(8\times8\times6)$ $k$-points (mean accuracy of $S_i J_{ij}^{\rm ex} S_j$ in the first-principles calculations) is sufficient for convergence.
Similar behavior can be seen in Fig.~\ref{fig:tc}(b), where $T_{\rm C}$ has been calculated
by a $9$-sublattice (i.e., $9$-inequivalent sites in Table~\ref{tab:parameter}) mean field analysis.\cite{herbst_magnetization_1984,skomski_finite-temperature_1998}
Compared with the MC results, the mean field results are less sensitive to $r_{\rm cut}$ and tend to overestimate $T_{\rm C}$.

%%%%%%%%%%%%%%%%%%%%%%%%%%%%%%%%%%%%%%%%%
%%%%%%     fig. magnetization curve
\begin{figure}%[b]
\includegraphics[width=8cm]{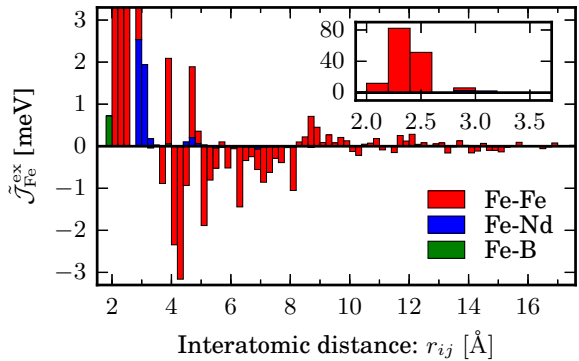}
\caption{\label{fig:jijsum}
Average Fe atom exchange coupling $\tilde{\mathcal{J}}^{\rm ex}_{\rm Fe}$ as a function of $r_{ij}$.
Inset shows a large-area view of $\tilde{\mathcal{J}}^{\rm ex}_{\rm Fe}$ in $r_{ij}<3.7\,\rm \AA$.
}
\end{figure}
%%%%%%%%%%%%%%%%%%%%%%%%%%
To analyze the long-range ($r_{ij} > 3.52\,{\rm \AA} $) exchange coupling effect for $T_{\rm C}$, Fig.~\ref{fig:jijsum} shows the average exchange coupling at the Fe atoms, $\tilde{\mathcal{J}}^{\rm ex}_{\rm Fe}(r_1,r_2)$, which is defined as follows:
\begin{eqnarray}
\tilde{\mathcal{J}}^{\rm ex}_{\rm Fe}(r_1,r_2)=\frac{1}{N_{\rm Fe}}\sum_{i \in {\rm Fe}, j} S_{i} J^{\rm ex}_{ij} S_j\ {\rm for}\ r_1<r_{ij} \leq r_2,
\label{eq:jsum}
\end{eqnarray}
where $N_{\rm Fe}$ is the total number of Fe sites.
Each bar height in Fig.~\ref{fig:jijsum} denotes the sum of $S_i J^{\rm ex}_{ij} S_j$ per atom in the range of each bar width (here $r_2-r_1=0.2\, {\rm \AA}$).
Because $\tilde{\mathcal{J}}_{\rm Fe}^{\rm ex}$ has many exchange bonds that correspond to a spherical surface area ($\propto r_{ij}^2$), it keeps small but significant value even in the long range. Indeed, the sum of short-range exchange couplings is $\tilde{\mathcal{J}}^{\rm ex}_{\rm Fe}(0,3.52\,{\rm \AA})=154.6\,$meV and that over a longer range is $\tilde{\mathcal{J}}^{\rm ex}_{\rm Fe}(3.52\,{\rm \AA},17.6\,{\rm \AA})=-10.2$ meV.
This negative value explains the decreasing trend for $T_{\rm C}$ shown in Fig.~\ref{fig:tc}.
The necessity of long-range exchange coupling has been identified for bcc-Fe \cite{spisak_theory_1997,singer_spin_2011,kvashnin_microscopic_2016} and MnBi,\cite{williams_extended_2016}
and so the dependence of $r_{\rm cut}$ appears to reflect the features of itinerant ferromagnetism.
Under the condition that $r_{\rm cut}=3.52$, $10.6$, and $17.6\,\rm \AA$,
each atom has approximately 13, 350, and 1660 exchange coupling bonds, respectively.
To reduce the computational load, we mainly consider $r_{\rm cut}=10.6\,\rm \AA$.
%which is sufficient for convergence of magnetic anisotropy. 

%$\langle A \rangle = \sum_n A({\bm R}_n) {\rm e}^{-E({\bm R}_n) k_B T} / \sum_n {\rm e}^{-E({\bm R}_n) k_B T }$,
%$n$ is Monte Carlo sampling, $E({\bm R}_n)$ and $A({\bm R}_n)$ are internal energy:$\,$$E$ and physical quantity:$\,$$A$ for spin configuration ${\bm R}_n$.

%av.200 exchange bond for 8.8, av.???? 15.8

%As mentioned
%stated above
%It is argued  that 
%need to consider long-range exchange coupling.
%itinerant ferromagnet

At the lower temperature point $T_{\rm sr}\,(\sim 145\,{\rm K})$ in Fig.~\ref{fig:magdis}, $\langle |M_z| \rangle$ reaches a maximum and $\red{M_{xy}=}\sqrt{\langle M_x^2 \rangle+\langle M_y^2 \rangle}$ approaches $0$, which is known to be the spin-reorientation transition of the Nd$_2$Fe$_{14}$B magnet.
The magnetization direction is tilted $34.4^\circ$ from the $z$-axis at $T=0$ for every $r_{\rm cut}$.
Above $T_{\rm sr}$, this direction exhibits uniaxial anisotropy along the $z$-axis.
In contrast to $T_{\rm C}$, $T_{\rm sr}$ has only a weak dependence on $r_{\rm cut}$.
The spin-reorientation transition is mainly driven by the higher-order terms ($l=4, 6$) of $A_l^{\red{m_l=0}}$ on the Nd atoms in Eq.~\eqref{eq:hami}.
Indeed, in comparison to the tilting angle of the single Nd atom at $T=0$ ($\theta=36.7^\circ$ in Fig.~\ref{fig:blm}(a)), we can see that the Fe magnetic anisotropy has little effect on the spin reorientation.
The reorientation property of Nd atoms is shared with the whole Nd$_2$Fe$_{14}$B through the exchange coupling $J^{\rm ex}_{\rm Fe \mathchar"712D Nd}$.
As shown in Fig.~\ref{fig:jijsum}, most contributions of $J^{\rm ex}_{\rm Fe \mathchar"712D Nd}$ are in the range $r_{\rm cut} \leq 3.4$\,\AA.
Therefore, $T_{\rm sr}$ has only a weak dependence on the long-range parts of $J^{\rm ex}_{ij}$.

%%%%%%%%%%%%%%%%%%%%%%%%%%%%%%%%%%%%%%%%
%%%%%%     fig. magnetization curve
\begin{figure}%[b]
\includegraphics[width=8cm]{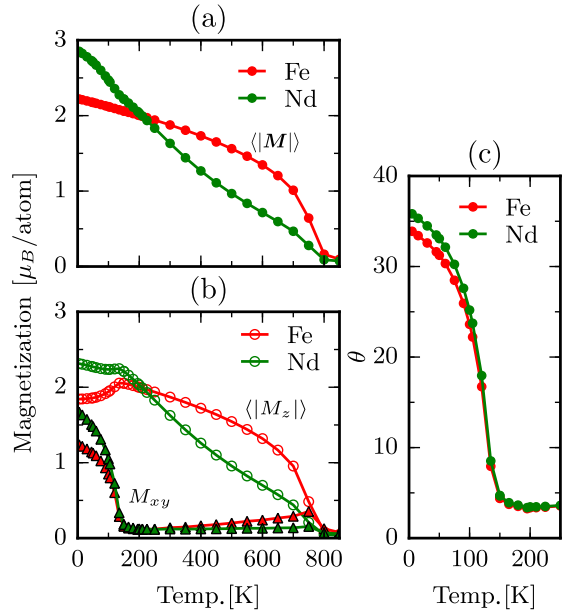}
\caption{\label{fig:magnd}
Temperature dependence of (a) the magnetization amplitude: $\langle|\bm{M}|\rangle$, (b) $z$-component: $\langle |M_{z}| \rangle$ and $xy$-component: $M_{xy}$ of Fe and Nd atoms for $r_{\rm cut}=10.6\,{\rm \AA}$ and $L=6$. (c) Each magnetization angle measured from $z$-axis at low-temperature region.
}
\end{figure}
%%%%%%%%%%%%%%%%%%%%%%%%%%
%
\red{
To look into the role for each atom in the above two transition at $T_{\rm C}$ and $T_{\rm sr}$, we plot in Fig.~\ref{fig:magnd} the temperature dependence of the magnetizations and the magnetization angle of Nd and Fe atoms.
In Fig.~\ref{fig:magnd}(a), \blue{reduction of the magnetization amplitude $\langle|M|\rangle$ with the temperature}
%the temperature decay of the magnetization amplitude, $\langle |\bm{M}|\rangle$, 
of each atom shows \blue{clear difference}.
%Here, we make mention of the difference of each magnetization curve in Fig.~\ref{fig:magnd}(a).
This difference is reflected by the amplitude of exchange couplings, 
$\tilde{\mathcal{J}}^{\rm ex}_{\rm Fe}$ ($\tilde{\mathcal{J}}^{\rm ex}_{\rm Nd}$) for $r_1=0$, $r_2=10.6\rm\,\AA$ is $142.9\,\rm meV$ ($33.5\,\rm meV$).
Hence, \blue{The ferromagnetic order of Fe is responsible to the magnetic order of the magnets.
At high temperature, we may have a picture that the magnetization of Nd atom is maintained by the interaction with the ordered Fe.}
%the ferromagnetic order of Fe atoms is a main cause of $T_{\rm C}$ and keeps the magnetization of Nd atoms at high-temperature region.
%
The rapidly decreasing of Nd magnetization with temperature corresponds to the poor thermal properties of magnetic anisotropy (see next section).
On the other hand, from each magnetization angle $\theta$ in Fig.~\ref{fig:magnd}(c), we can verify that $T_{\rm sr}$ is mainly depend on the magnetic anisotropy of Nd atoms as was mentioned in the previous paragraph.
The magnetization angle $\theta$ is calculated by using $\langle|M_z|\rangle$ and $M_{xy}$ in Fig.~\ref{fig:magnd}(b) as follows:
\begin{eqnarray}
\theta=\arctan{\left( \frac{ \langle |M_z| \rangle}{M_{xy}} \right)}. 
%sqrt{\langle M_x^2 \rangle+\langle M_y^2 \rangle} }\right)}
\end{eqnarray}
%
%which is calculated from the magnetization of each atom (Fig.~\ref{fig:magnd} (a)).
%
In Fig.~\ref{fig:magnd}(c), the angle of Nd magnetization has always larger value than the angle of Fe magnetization below $T_{\rm sr}$.
This behavior implies that the spin-reorientation occurs because that the tilted Nd magnetization attracts the Fe magnetization.
}

\red{
It is necessary to keep in mind that the model parameters do not include the thermal variations of the lattice parameters and the electronic states.}
\blue{However,} despite using many parameters from first-principles calculations, the above thermodynamic results ($T_{\rm C}\sim 754\,\rm K$, $T_{\rm sr}\sim 145\,\rm K$ for $r_{\rm cut}=10.6\,\rm\AA$) are \red{basically} consistent with experimental values ($T_{\rm C}\sim 585\,\rm K$, $T_{\rm sr}\sim 135\,\rm K$).\cite{herbst_ndfe_14b_1991}
Therefore, the model and the parameter sets are sufficiently reliable for studying the temperature dependence of magnetic anisotropy in Nd$_2$Fe$_{14}$B.

%
%%%%%%%%%%%%%%%%%%%%%%%%%%%%%%%%%%%%%%%%%%%%%%%%%%%%%

\subsection{Temperature Dependence of Magnetic Anisotropy}\label{sec:aniso}
We now discuss the temperature dependence of magnetic anisotropy.
%
%
%%%%%%     fig. fit torq for each model
\begin{figure}%[b]
\includegraphics[width=8cm]{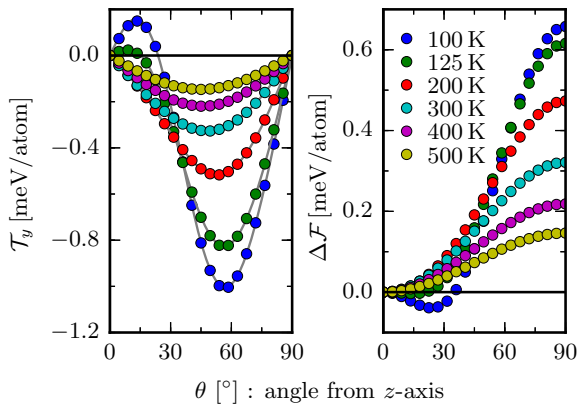}% Here is how to import EPS art
\caption{\label{fig:torq}
Angular dependence of $y$-direction torque $\mathcal{T}_y$ (left side) and free energy $\Delta \mathcal{F}$ (right side) at each temperature for $r_{\rm cut}=10.6\,{\rm \AA}$ and $L=4$. 
The gray lines on the left side show the fit of the torque data to $-\partial \Delta\mathcal{F}/\partial \theta$ in Eq.~\eqref{eq:fitting}.
}
\end{figure}
%%%%%%%%%%%%%%%%%%%%%%%%%%%
%
Figure~\ref{fig:torq} shows the $y$-direction torque $\mathcal{T}_y$ and free energy $\Delta \mathcal{F}$ as a function of magnetization angle $\theta$ for $L=4$ as calculated by the C-MC method.
In the present paper, the directions of magnetization constrained by the C-MC method are rotated by $\theta$ around the $y$-axis.
Therefore, the torque is perpendicular to the $x$-$z$ plane, i.e., both the $x$ and $z$ components of torque are zero.
%

%%%%%%     fig. Ha for each model
\begin{figure}%[b]
\includegraphics[width=8cm]{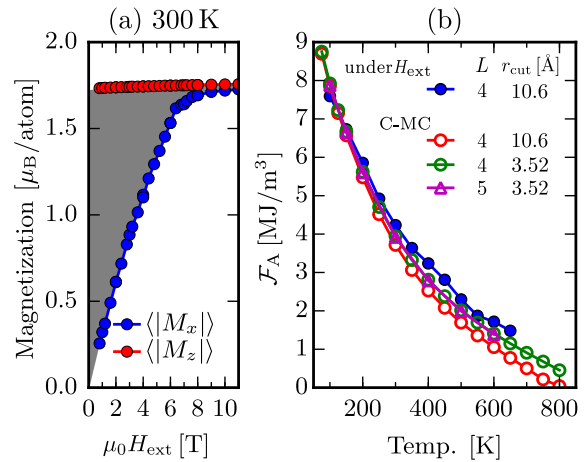}% Here is how to import EPS art
\caption{\label{fig:ha}
(left) Magnetization curve $\langle |M_{z(x)}| \rangle$ under the $z(x)$-direction of external magnetic field $H_{\rm ext}$;
gray area corresponds to magnetic anisotropy energy $\mathcal{F}^H_{\rm A}$ in the blue solid lines on the right-hand panel. 
(right) $\mathcal{F}^{(H)}_{\rm A}$ for the four calculation conditions.}
\end{figure}
%%%%%%%%%%%%%%%%%%%%%%%%%%%
%
To verify the C-MC method, we compare the magnetic anisotropy energies $\mathcal{F}_{\rm A}$ with those given by the typical MC method, $\mathcal{F}^{H}_{\rm A}$.
Here, $\mathcal{F}_{\rm A}$ is defined as $\Delta \mathcal{F}_{\rm max}-\Delta\mathcal{F}_{\rm min}$ in Fig.~\ref{fig:torq}, and $\mathcal{F}^H_{\rm A}$ is derived from the magnetization curves as the gray area on the left of Fig.~\ref{fig:ha} (example at $T=300\,\rm K$),
where $\langle |M_{x (z)}| \rangle$ is the magnetization curve under $H_{\rm ext}$ in the $x (z)$-direction.
From the right of Fig.~\ref{fig:ha},
we can confirm that $\mathcal{F}_{\rm A}$ is in good agreement with $\mathcal{F}^{H}_{\rm A}$, particularly in the low-temperature region, although $\mathcal{F}^{H}_{\rm A}$ tends to give an overestimate.
This overestimate occurs because, at finite temperatures, the effective magnetic anisotropy of each spin decreases as a result of thermal fluctuations.
When evaluating $\mathcal{F}^{H}_{\rm A}$, the thermal fluctuations are suppressed by the external field to saturate the magnetization.
This suppression becomes stronger as the temperature increases,
causing the overestimation to be significant in high-temperature region.

\begin{figure}%[b]
\includegraphics[width=8cm]{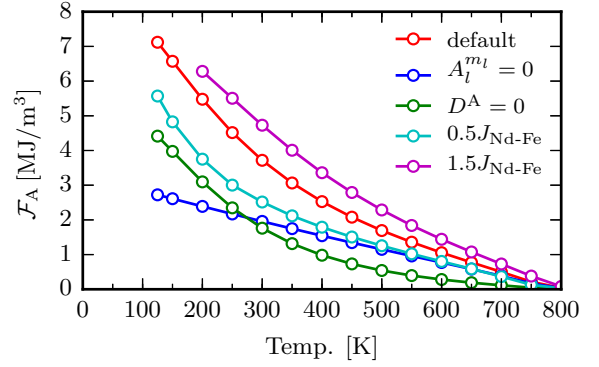}% Here is how to import EPS art
\caption{\label{fig:hikaku} 
Comparison of the temperature dependence of magnetic anisotropy energy $\mathcal{F}_{\rm A}$ in five cases (details in text) for $r_{\rm cut}=10.6\,\rm \AA$ and  $L=4$.}
\end{figure}

We also plot $\mathcal{F}_{\rm A}$ for other calculation conditions: $(L,\,r_{\rm cut})=$  $(4,\,3.52)$ and $(5,\,3.52)$ on the right of Fig.~\ref{fig:ha}.
These results show that a system size of $L=4$ is sufficient to obtain convergence in the magnetic anisotropy.
Additionally, the length of $r_{\rm cut}$ affects $\mathcal{F}_{\rm A}$ at high temperatures.
As mentioned in terms of spin reorientation,
%the magnetic anisotropy of Nd is essentially unaffected by differences in $r_{\rm cut}$, but there is a direct effect on the anisotropy of Fe.
%
%Therefore, in temperature regions where the anisotropy of Fe becomes dominant (i.e., $T\geq 250\,\rm K$, see Fig.~\ref{fig:hikaku}),
%the effects on $\mathcal{F}_{\rm A}$ of differences in $r_{\rm cut}$ are clearly evident. 
%
\red{the magnetic anisotropy of Nd is essentially unaffected by differences in $r_{\rm cut}$.
Hence, it can be regarded as that the difference between red and green lines in Fig.~\ref{fig:ha}(b) occurs due to $r_{\rm cut}$ dependence of Fe anisotropy.
%difference between $\mathcal{F}_{\rm A}$ for $(L,\,r_{\rm cut})=$  $(4,\,3.52)$ and $(4,\,10.6)$ in Fig.~\ref{fig:ha}(b)
}
Therefore, \red{in high-temperature \blue{region} where Fe anisotropy becomes larger than the Nd anisotropy} (see Fig.~\ref{fig:hikaku} \red{$A_l^{m_l}=0$ and $D^{\rm A}=0$}), the effects on $\mathcal{F}_{\rm A}$ of differences in $r_{\rm cut}$ are clearly evident.

Returning to Fig.~\ref{fig:torq},
we can see that for $100\,\rm K$ and $125\,\rm K$, the torque (free energy) curve attains a local maximum (minimum) at $\theta\neq 0$, which reflect the spin reorientation (shown in Fig.~\ref{fig:magdis}).
In contrast, above $T \geq 200\,\rm K$, the local maximum (minimum) disappears and the torque (free energy) curve approaches $\propto \sin 2\theta$ ($\sin^2 \theta$).
This behavior implies that the magnetic anisotropy constant $K^{\rm A}_1$ becomes dominant as the temperature increases.
%
%%%%%%     fig. Ku for each models
\begin{figure}%\mathcal{F}_[b]
\includegraphics[width=8cm]{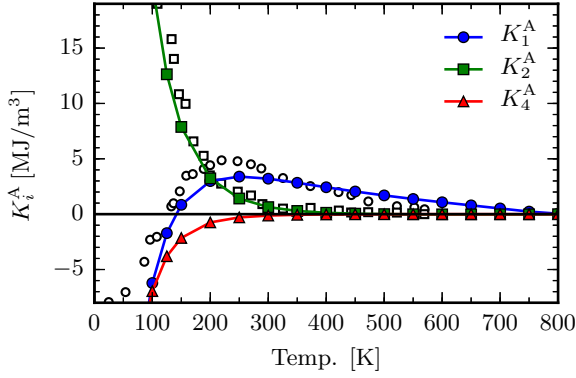}% Here is how to import EPS art
\caption{\label{fig:ku} Anisotropy constants $K^{\rm A}_m$ as a function of temperature for $r_{\rm cut}=10.6\,\rm\AA$ and $L=4$.
White circles and squares indicate experimental results\cite{durst_determination_1986} for $K^{\rm A}_{1}$ and $K^{\rm A}_{2}$, respectively.
}
\end{figure}
%%%%%%%%%%%%%%%%%%%%%%%%%%%
%
%
To clarify the temperature dependence, Fig.~\ref{fig:ku} shows the magnetic anisotropy constants $K^{\rm A}_m\ (m=1,2,4)$ that were calculated by fitting $\mathcal{T}_y$ in Fig.~\ref{fig:torq} to the torque equation:
%
%%%%%%%%%%%%%%%%%%%%%%%%%%%
\begin{eqnarray}
	\mathcal{T}_y(\theta,T) &=& - \frac{\partial}{\partial \theta} \Delta\mathcal{F}(\theta,T), \nonumber \\
	\Delta\mathcal{F}(\theta,T)&=&K^{\rm A}_1(T) \sin^2\theta + K^{\rm A}_2(T) \sin^4\theta + K^{\rm A}_4(T) \sin^6\theta. \nonumber \\
%	
%	\mathcal{T}_y(\theta,T) &=& - \frac{\partial \mathcal{F}_{\rm A}(\theta,T)}{\partial \theta} \nonumber \\
%					   		&=& -\sin 2\theta \left[ K^{\rm A}_1(T) + 2K^{\rm A}_2(T) \sin^2\theta \right. \nonumber \\
%					   		& & \qquad + \left. 3K^{\rm A}_3(T)\sin^4\theta\right].
	\label{eq:fitting}
\end{eqnarray}
%%%%%%%%%%%%%%%%%%%%%%%%%%%
%
These constants can only be calculated correctly using the C-MC method.
%
%K^{\rm A}
We can confirm that $K^{\rm A}_2$ and $K^{\rm A}_4$ tend to zero and $K^{\rm A}_1$ becomes dominant in the region of $T>300\rm\, K$.
Additionally, $K^{\rm A}_1$ becomes negative in the low-temperature region. This is reflected by the local minimum of $\Delta \mathcal{F}$ in Fig.~\ref{fig:torq}, indicating the spin reorientation transition.
The temperature dependence of $K^{\rm A}_m$ agrees reasonably well with previous experimental results\cite{hirosawa_magnetization_1986,yamada_magnetocrystalline_1986,durst_determination_1986} and mean field theory.\cite{sasaki_theoretical_2015,miura_direct_2015}
Note that, at $T<100\,\rm K$, all of the $|K^{\rm A}_m|$ are significantly larger than the experimental values.
For classical spin systems, this deviation in $K^{\rm A}_m$ (and also $M$) is finite at zero temperature on account of the infinite degrees of freedom of classical spin (for quantum spin systems, the deviations of $K^{\rm A}_m$ and $M$ at $T=0$ are zero).\cite{miura_direct_2015}
This explain the difference between our results and the experimental results at $T<100\,\rm K$.

%%%%%%     fig. 
%%%%%%%%%%%%%%%%%%%%%%%%%%%

To examine the relationship between the exchange coupling and magnetic anisotropy,
we considered various input parameter sets.
Figure~\ref{fig:hikaku} shows the anisotropy energy $\mathcal{F}_{\rm A}$ for five cases:
the same result as shown by the red lines in Fig.~\ref{fig:ha} (default),
a model including only Fe magnetic anisotropy ($A_l^{\red{m_l}}=0$),
a model including only Nd magnetic anisotropy ($D^{\rm A}=0$),
a model with all $J_{\rm Fe \mathchar"712D Nd}^{\rm ex}$ reduced by half (0.5$J_{\rm Fe \mathchar"712D Nd}^{\rm ex}$),
and a model with all $J_{\rm Fe \mathchar"712D Nd}^{\rm ex}$ increased by half (1.5$J_{\rm Fe \mathchar"712D Nd}^{\rm ex}$).
In the case of $A_l^{\red{m_l}}=0$, the anisotropy energy decreases almost linearly with temperature.
This behavior is a typical property of the classical Heisenberg models that include only $\sin^2 \theta$ for the anisotropy energy.
%
%implying the relational expression of $K^{\rm A}_1(T) \propto M(T)^3$ based on Callen-Callen law\cite{callen_static_1963,callen_present_1966} (however actually, the case of ($A_l^m=0$) deviate from the law, detail in Fig.~\ref{fig:callen}).
%
In contrast, the case of $D^{\rm A}=0$ exhibits a rapid decrease, which can be explained by the difference in the exchange coupling $\tilde{\mathcal{J}}^{\rm ex}_{\rm atom}(r_1,r_2)$ of Nd and Fe atoms (see Eq.~\eqref{eq:jsum}). 
We have that $\tilde{\mathcal{J}}^{\rm ex}_{\rm Fe}$ and $\tilde{\mathcal{J}}^{\rm ex}_{\rm Nd}$ for $r_1=0$, $r_2=10.6\rm\,\AA$ are $142.9\,\rm meV$ and $33.5\,\rm meV$, respectively.
%
%Now, $\tilde{\mathcal{J}}^{\rm ex}_{\rm atom}(r_1,r_2)$ of Fe and Nd atoms for $r_1=0$, $r_2=10.6\rm\,\AA$ are $\tilde{\mathcal{J}}^{\rm ex}_{\rm Fe}=142.9\,\rm meV$ and $\tilde{\mathcal{J}}^{\rm ex}_{\rm Nd}=33.5\,\rm meV$, respectively.
%
Here, $\tilde{\mathcal{J}}^{\rm ex}_{\rm Fe (Nd)}$ is almost given by the Fe-Fe (Nd-Fe) exchange couplings (see Fig.~\ref{fig:jex}).
The Nd atoms, which give the whole Nd$_2$Fe$_{14}$B system magnetic anisotropy through $\tilde{\mathcal{J}}^{\rm ex}_{\rm Nd}$, are highly susceptible to thermal fluctuations,
unlike the Fe atoms, which play a key role in magnetism (such as $|\bm{M}|$ and $T_{\rm C}$).
This difference in thermal susceptibility explains the rapid decrease in $\mathcal{F}_{\rm A}$ for $D^{\rm A}=0$.
For the same reason, in the case of 0.5$J_{\rm Fe \mathchar"712D Nd}^{\rm ex}$, which includes both $A_l^{\red{m_l}}$ and $D^{\rm A}$,
$\mathcal{F}_{\rm A}$ decreases rapidly with temperature, and approaches $A_l^{\red{m_l}}=0$ at approximately $400\,\rm K$.
This means that the effects of Nd magnetic anisotropy are almost wiped out by thermal fluctuations above $400\,\rm K$.
However, for 1.5$J_{\rm Fe \mathchar"712D Nd}^{\rm ex}$, $\mathcal{F}_{\rm A}$ is almost linear.
The above discussion for Fig.~\ref{fig:hikaku} allows us to understand that 
$\tilde{\mathcal{J}}^{\rm ex}_{\rm Nd}$ (rather than $\tilde{\mathcal{J}}^{\rm ex}_{\rm Fe}$)
makes a strong contribution to the magnetic anisotropy of Nd atoms,
which supports the results of previous studies.\cite{sasaki_theoretical_2015,matsumoto_relevance_2016}
%
%$\tilde{\mathcal{J}}^{\rm ex}_{\rm Nd}$ strongly affect on the magnetic anisotropy,
%supporting the previous study,\cite{sasaki_theoretical_2015,matsumoto_relevance_2016}
%%
%Additionally, the magnetic anisotropy of Nd atoms does not relate to $\tilde{\mathcal{J}}^{\rm ex}_{\rm Fe}$, namely $T_{\rm C}$.
%

%%%%%%     fig. 
\begin{figure}%[b]
\includegraphics[width=8cm]{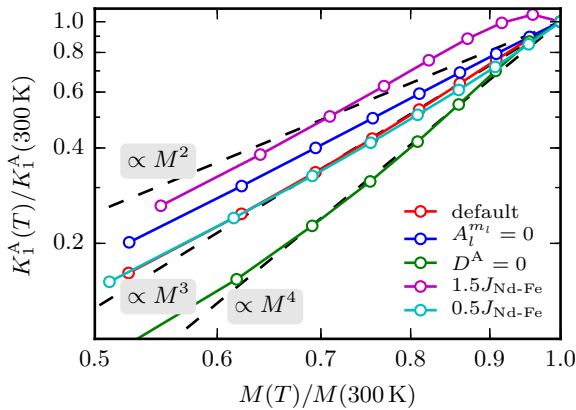}% Here is how to import EPS art
\caption{\label{fig:callen}
Relation between $K^{\rm A}_1(T)$ and $M(T)$ at each temperature for the same parameter sets and calculation conditions in Fig.~\ref{fig:hikaku}.
The natural logarithm is taken for both axes.
The Callen--Callen law corresponds to $M(T)^3$, illustrated by a dashed line.}
\end{figure}
%%%%%%%%%%%%%%%%%%%%%%%%%%%

To analyze the results shown in Fig.~\ref{fig:hikaku} in the context of the Callen--Callen law,\cite{callen_static_1963,callen_present_1966}
i.e., $K^{\rm A}_1(T) \propto M(T)^3$ for $K^{\rm A}_2=K^{\rm A}_4=0$, 
Fig.~\ref{fig:callen} illustrates the relationship between $K^{\rm A}_1$ and $M$ above $300\,\rm K$.
It is clear that 1.5$J_{\rm Fe \mathchar"712D Nd}^{\rm ex}$ deviates from this law, because $K^{\rm A}_2$ is comparable to $K^{\rm A}_1$ at $300\,\rm K$.
Varying the anisotropy terms $A_l^{\red{m_l}}$ and $D^{\rm A}$ affects these relations more than $\tilde{\mathcal{J}}^{\rm ex}_{\rm Nd}$.
For $A_l^{\red{m_l}}=0$, Nd magnetization decreases rapidly with temperature, whereas Fe anisotropy decreases gradually.
Hence, $K^{\rm A}_1/M$ tends to increase.
Conversely, for $D^{\rm A}=0$, Fe magnetization slowly decreases with temperature whereas Nd anisotropy decreases rapidly, hence $K^{\rm A}_1/M$ tends to decrease.
As above two effects happen to cancel out, the default case and 0.5$J_{\rm Fe \mathchar"712D Nd}^{\rm ex}$ agree with the Callen--Callen law.

%

%implying the relational expression of $K^{\rm A}_1(T) \propto M(T)^3$ based on Callen-Callen law\cite{callen_static_1963,callen_present_1966} (however actually, the case of ($A_l^m=0$) deviate from the law, detail in Fig.~\ref{fig:callen}).
%\sout{
%It is noted that linearly decreasing of $\mathcal{F}_{\rm A}$ for ($A_l^m=0$) in Fig~\ref{fig:hikaku} generally imply the Callen-Callen law,
%on the contrary, it deviate from the law seen in Fig.~\ref{fig:callen}.
%%
%In this case, since Nd atoms do not have anisotropy and do not strongly interact with Fe atoms, it does not appear to $\mathcal{F}_{\rm A}$ that the effects of the rapidly decreasing Nd magnetization on the anisotropy.
%%%%	
%%Nd (and also B) magnetization rapidly decrease with temperature compared with Fe, in other words,
%%$\tilde{\mathcal{J}}^{\rm ex}_{\rm Nd}$ is much smaller than $\tilde{\mathcal{J}}^{\rm ex}_{\rm Fe}$.
%%
%The magnetic inhomogeneity between Fe and Nd cause the above discrepancy.
%}

%This inhomogeneity of exchange couplings, namely $\tilde{\mathcal{J}}^{\rm ex}_{\rm Nd}$ is much smaller than $\tilde{\mathcal{J}}^{\rm ex}_{\rm Fe}$,

The Callen--Callen law was derived under the assumption of homogeneous ferromagnetic and single-ion anisotropy systems at temperatures far from $T_{\rm C}$.
\red{
Therefore, it is natural that multi-sublattice model such as Nd$_2$Fe$_{14}$B does not follow the Callen--Callen law, which was also pointed out by using a mean field approach.\cite{skomski_finite-temperature_2006}
}
Additionally, in actual ferromagnetic metals that have two-ion magnetic anisotropy, 
the temperature dependence of magnetic anisotropy deviates from Callen--Callen law,\cite{thiele_temperature_2002,okamoto_chemical-order-dependent_2002,staunton_temperature_2006,asselin_constrained_2010,kobayashi_effects_2016}
such as $L1_0$-FePt, $K^{\rm A}_1(T) \propto M^{2.1}(T)$.\cite{okamoto_chemical-order-dependent_2002}
Therefore, more detailed discussion of the temperature dependence is needed to formulate the theory for itinerant electrons and inhomogeneous systems.

%
%The Callen--Callen law was derived under the assumption of homogeneous ferromagnetic and single-ion anisotropy systems at temperatures far from $T_{\rm C}$.
%In actual ferromagnetic metals that have two-ion magnetic anisotropy, such as Fe-based systems, 
%the temperature dependence of magnetic anisotropy is expected to follow $K^{\rm A}_1(T) \propto M^2(T)$.\cite{thiele_temperature_2002,okamoto_chemical-order-dependent_2002,staunton_temperature_2006,asselin_constrained_2010,kobayashi_effects_2016}
%Therefore, more detailed discussion of the temperature dependence is needed to formulate the theory for itinerant electrons and inhomogeneous systems.
%

%This stems from the situation that the exchange couplings between Nd moments and Fe spins are much smaller than those between Fe spins.
%Callen-Callen law is only valid for homogeneous and single ion anisotropy systems at temperatures far from the $T_{\rm C}$.

%For (default, 0.5, 1.5 ), corresponding to Callen-Callen law is coincidence.

\subsection{Energy Barrier}
Finally, we discuss the external magnetic field $H_{\rm ext}$ response of the energy barrier (activation energy)\cite{gaunt_magnetic_1986,chantrell_calculations_2000,suess_reliability_2007,bance_thermally_2015,bance_thermal_2015} which governs the probability of magnetization reversal via the thermal fluctuation of spins. 
If this response can be measured experimentally,\cite{goto_energy_2015} it would allow the magnetic coercivity mechanism to be predicted at finite temperatures.
%
%%%%%%     fig. 
\begin{figure}[t]
\includegraphics[width=8cm]{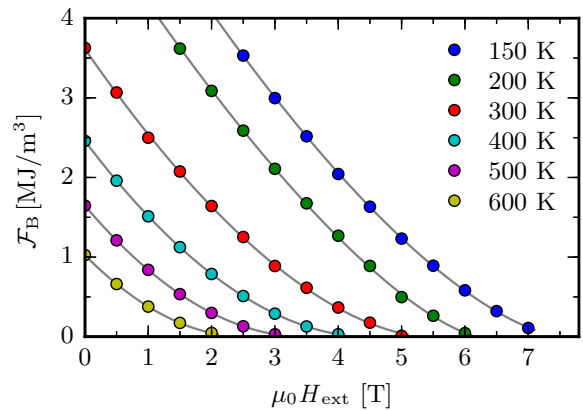}% Here is how to import EPS art
\caption{\label{fig:barrier}
Height of the energy barrier, $\mathcal{F}_{\rm B}$, as a function of external magnetic field, $H_{\rm ext}$, at each temperature
for $r_{\rm cut}=10.6\,\rm\AA$ and $L=4$.
The gray lines illustrate the fit to Eq.~\eqref{eq:viscosity}.
}
\end{figure}
%%%%%%%%%%%%%%%%%%%%%%%%%%%
%
Figure~\ref{fig:barrier} shows the height of the energy barrier, $\mathcal{F}_{\rm B}$, when $H_{\rm ext}$ is applied opposite to the $z$-direction of $\bm{M}$.
By uniformly rotating the direction of $\bm M$ using the C-MC method, we evaluated $\mathcal{F}_{\rm B}$;
therefore, $\mathcal{F}_{\rm B}=\mathcal{F}_{\rm A}$ for $H_{\rm ext}=0$.
The $H_{\rm ext}$ response of $\mathcal{F}_{\rm B}$ is generally expressed by:\cite{gaunt_magnetic_1986}
\begin{eqnarray}
\mathcal{F}_{\rm B}(H_{\rm ext}) = \mathcal{F}^0_{\rm B}(1-H_{\rm ext}/H_0)^n,
\label{eq:viscosity}
\end{eqnarray}
where $\mathcal{F}^0_{\rm B} = \mathcal{F}_{\rm B}(H_{\rm ext}=0) = \mathcal{F}_{\rm A} $, and $H_0$ is equal to the value of $H_{\rm ext}$ at $\mathcal{F}_{\rm B}=0$, which corresponds to the upper limit of the coercive field, $H_c$, under uniform rotation.
For finite temperatures, \sout{\red{ and non-uniform rotation,}}
the thermal fluctuation helps the magnetization reversal to overcome the energy barrier, and so $H_c$ is much lower than $H_0$.
The exponent $n$ can take various values, such as $n=2$ for the Stoner--Wohlfarth model and $n=1$ for the weak domain-wall pinning mechanism.\cite{gaunt_magnetic_1986}

\begin{figure}%[t]
\includegraphics[width=8cm]{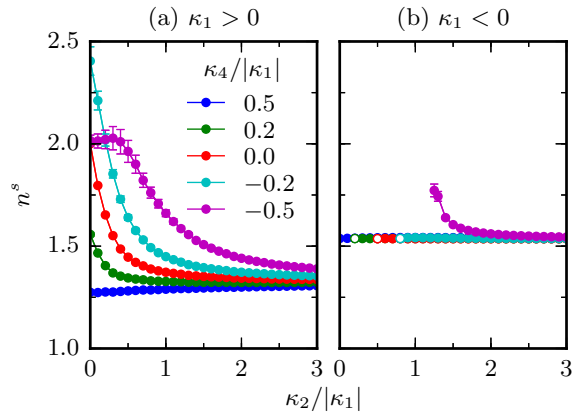}% Here is how to import EPS art
\caption{\label{fig:n}
The exponent $n^s$ in the magnetic field response for the single-spin model, Eq.~\eqref{eq:single}, as a function of $\kappa_2/|\kappa_1|\,(\geq 0)$ with fitting error bars ($95\,\%$ confidence) on each $\kappa_4/|\kappa_1|$ for (a) $\kappa_1>0$ and (b) $\kappa_1<0$.
Energy barrier $\mathcal{F}_B$ disappears when $\kappa_2/|\kappa_1|$ is below each white point in (b).
}
\end{figure}

%%%%%%%%%%%%%%%%%%%%%
\begin{table}%[b]
\caption{\label{tab:n} Fitting parameters ($\mathcal{F}^0_{\rm B}$, $H_0$, $n$) at each temperature.
The exponent $n^s_{K}$ was estimated with the single-spin model (Eq.~\eqref{eq:single})
using the anisotropy constant $K^{\rm A}_m$ in Fig.~\ref{fig:ku} instead of $\kappa_m$.
}
\begin{ruledtabular} 
\begin{tabular}{cccccccc}
Temp. [$\rm K$] & $\mathcal{F}^0_{\rm B}\,[\rm MJ/m^3]$ & $\mu_0 H_0\,[\rm T]$ & $n$ & $\frac{K^{\rm A}_2}{K^{\rm A}_1}$& $\frac{K^{\rm A}_4}{K^{\rm A}_1}$ &$n^s_K$\\
\hline
150  &  6.53 & 7.54  & 1.53 &  9.35 & -2.55 & 1.56\\
200  &  5.37 & 6.23  & 1.42 &  1.08 & -0.25 & 1.44\\
300  &  3.61 & 5.41  & 1.72 &  0.2  & -0.04 & 1.72\\
400  &  2.46 & 4.44  & 1.90 &  0.05 & -0.01 & 1.91\\
500  &  1.65 & 3.46  & 1.97 &  0    &  0    & 2.00\\
600  &  1.03 & 2.55  & 2.00 & -0.02 &  0    & 2.05\\
\end{tabular}
\end{ruledtabular}
%\footnotetext[1]{We}
\end{table}
%%%%%%%%%%%%%%%%%%%%%%%%%%%
%
%
The parameters $\mathcal{F}^0_{\rm B}$, $H_0$, and $n$ were obtained by fitting $\mathcal{F}_{\rm B}(H_{\rm ext})$ in Fig.~\ref{fig:barrier}, and are listed in Table~\ref{tab:n}.
We can see that $n$ takes values of less than $2$ in the low-temperature region (below the room temperature, $T_{\rm R}
\sim 300\,\rm K$) and approaches $2$ as the temperature increases.
This reflects the fact that the magnetic anisotropy is mainly governed by the $K^{\rm A}_1$ term in the high-temperature region (see Fig.~\ref{fig:ku}).
To clarify this, we estimated the exponent $n^s$ by fitting from the anisotropy energy of the single-spin model, which is defined as:
\begin{eqnarray}
E^s_{\rm A}(\theta)=\kappa_1 \sin^2 \theta + \kappa_2 \sin^4 \theta + \kappa_4 \sin^6 \theta + mH_{\rm ext}\cos\theta\red{.} \nonumber \\
\label{eq:single}
\end{eqnarray}
With $\kappa_2=\kappa_4=0$, this corresponds to the Stoner--Wohlfarth model.
The dependence of the anisotropy constant on $n^s$ is plotted in Fig.~\ref{fig:n}.
This figure confirms that $\kappa_2$ and $\kappa_4$ have a significant effect on $n^s$ for (a) $\kappa_1>0$,
whereas $n^s$ is less sensitive for (b) $\kappa_1<0$, which corresponds to the low-temperature region below \red{$T_{\rm sr}$} of Nd$_{2}$Fe$_{14}$B (see Fig.~\ref{fig:ku}).
Here, the deviation of $n^s$ given by fitting Eq.~\eqref{eq:viscosity} becomes large when either $|\kappa_2|$ or $|\kappa_4|$ increases.
Therefore, near the points where fitting error bars are large (see Fig.~\ref{fig:n}),
we should pay attention to the $n^{s}$ values, which are dependent on fitting procedures.

%%%%%%     fig. 
%%%%%%%%%%%%%%%%%%%%%%%%%%%

Additionally, we input $K^{\rm A}_m$ (from Fig.~\ref{fig:ku}) into $\kappa_m$ in Eq.~\eqref{eq:single},
and estimated the exponent $n^s_K$ listed in Table \ref{tab:n}.
Despite using the single-spin model, $n^s_K$ is in good agreement with $n$, where $n$ has been evaluated on an inhomogeneous spin system such as Nd$_2$Fe$_{14}$B.
This indicates that, in  terms of the magnetic field response of uniform rotation,
the anisotropy constants $K^{\rm A}_m$ are renormalized by the magnetic inhomogeneities and thermal fluctuations.
For the Nd$_2$Fe$_{14}$B system, in particular, we should bear in mind that the response occurs for $n<2$
when below room temperature, $T_{\rm R}$.
\\


\nocite{*}

\section{Summary}
%We investigated and discussed the magnetic properties at finite temperature of the Nd$_2$Fe$_{14}$B bulk system,
%by applying constrained Monte Carlo method to the realistic classical three dimensional Heisenberg model with many first principles calculations data.

We have constructed a realistic classical three-dimensional Heisenberg model using parameters from first-principles calculations,
and investigated the magnetic properties of the Nd$_2$Fe$_{14}$B bulk system at finite temperatures.
Applying the constrained Monte Carlo method to this model,
from atomic-scale parameters,
we evaluated macroscopic magnetic anisotropies which include correctly magnetic inhomogeneities and thermal fluctuations.
Despite using many parameters from first-principles calculations (except for $A_l^{\red{m_l}}$),
the model reproduced the experimentally observed spin reorientation and magnetic anisotropy constants $K^{\rm A}_m$.

Using this calculation system, we found that,
because the exchange couplings between Nd moments and Fe spins are much smaller than those between Fe spins,
the magnetic anisotropy of Nd atoms decreases more rapidly than that of Fe atoms.
Additionally, owing to this magnetic inhomogeneity, the temperature dependence of $K^{\rm A}_1$ deviates from the Callen--Callen law, even above room temperature ($T_{\rm R}\sim300\rm\,K$), when the Fe (Nd) anisotropy terms are removed to leave only the Nd (Fe) anisotropy.
Furthermore, we also found that the exponent $n$ in the magnetic field response of barrier height is less than $2$ in the low-temperature region below $T_{\rm R}$, whereas $n$ approaches $2$ when $T>T_{\rm R}$, indicating Stoner--Wohlfarth-type magnetization rotation.
This behavior reflects the fact that the magnetic anisotropy is mainly governed by the $K^{\rm A}_1$ term in $T>T_{\rm R}$,
which is explained by the single-spin model with a renormalized $K^{\rm A}_m$.

We have a plan to extend the constructed framework in \red{present} paper to non-uniform magnetization reversal in finite-size particles, including the effects of the grain surfaces or grain boundaries.

%\\
%%%%%%%%%%%%%%%%%%%%%%%%%%%%%%%%%%%%%%

\begin{acknowledgments}
We would like to thank D.~Miura, R.~Sasaki, M.~Nishino, Y.~Miura, and S.~Hirosawa for useful discussions and information.
This work is supported by the Elements Strategy Initiative Project under the auspices of MEXT.

%Some of the numerical computations were carried out at the Cyberscience Center, Tohoku University.

%Part of the experimental results in this research were obtained using supercomputing resources at Cyberscience Center, Tohoku University.

%This work is supported by a Grant-in-Aid for Scientific Research (C) and also by the Next Generation Supercomputing Project, Nanoscience Program from MEXT of Japan.
\end{acknowledgments}

\bibliographystyle{apsrev4-1} % Tell bibtex which bibliography style to use
\bibliography{05.ndfeb_spin01}  % Tell bibtex which .bib file to use (this one is some example file in TexLive's file tree)

%\begin{references}
%

%\end{references}

\end{document}